\documentclass[9pt,letterpaper,twocolumn]{extarticle}
\usepackage{stmaryrd}
\usepackage{graphicx}
\usepackage{wrapfig}
\usepackage[latin1]{inputenc}
\usepackage[font=small,labelfont={sf,bf},labelsep=period]{caption}
\usepackage{setspace}
\usepackage[hypertexnames=false,plainpages=false,pdfpagelabels,colorlinks=true,linkcolor=blue,urlcolor=blue,citecolor=blue,pdftitle={Negative Absolute Temperature for Motional Degrees of Freedom},pdfauthor={Simon Braun},pdfdisplaydoctitle=true,pdfpagelayout=OneColumn,pdfduplex=DuplexFlipLongEdge]{hyperref}

\newcommand{\mboxs}[1]{\mbox{\scriptsize{#1}}}
\newcommand{\vspacelarge}{0.6cm}
\newcommand{\vspacesmall}{0.1cm}
\newcommand{\vspacereferences}{-0.6cm}

\begin{document}

\singlespacing
\noindent
{\center{\textsf{\textbf{\Huge{Negative Absolute Temperature for Motional Degrees of Freedom\\}}}}}
\singlespacing

\vspace{-0.1cm}

\noindent
{\center{\textsf{\large{\textbf{S. Braun,$^{1,2}$ J. P. Ronzheimer,$^{1,2}$ M. Schreiber,$^{1,2}$\\S. S. Hodgman,$^{1,2}$ T. Rom,$^{1,2}$ I. Bloch,$^{1,2}$\\U. Schneider$^{1,2*}$\\}}}}}

\vspace{0.2cm}

\noindent
\textsf{$^1$ Fakultät für Physik, Ludwig-Maximilians-Universit\"at M\"un\-chen, Schellingstr.\ 4, 80799 Munich, Germany}

\vspace{0.1cm}

\noindent
\textsf{$^2$ Max-Planck-Institut f\"ur Quantenoptik, Hans-Kopfermann-Str.\ 1, 85748 Garching, Germany}

\vspace{0.1cm}

\noindent
\textsf{$^*$ E-mail: ulrich.schneider@lmu.de}


\vspace{0.6cm}

\noindent
\textsf{Absolute temperature, the fundamental temperature scale in thermodynamics, is usually bound to be positive. Under special conditions, however, negative temperatures - where high-energy states are more occupied than low-energy states - are also possible. So far, such states have been demonstrated in localized systems with finite, discrete spectra. Here, we were able to prepare a negative temperature state for motional degrees of freedom. By tailoring the Bose-Hubbard Hamiltonian we created an attractively interacting ensemble of ultracold bosons at negative temperature that is stable against collapse for arbitrary atom numbers. The quasi-momentum distribution develops sharp peaks at the upper band edge, revealing thermal equilibrium and bosonic coherence over several lattice sites. Negative temperatures imply negative pressures and open up new parameter regimes for cold atoms, enabling fundamentally new many-body states and counterintuitive effects such as Carnot engines above unity efficiency.}


\vspace{0.4cm}

\begin{figure}[h!t]
	\centering
	\includegraphics[width=84mm]{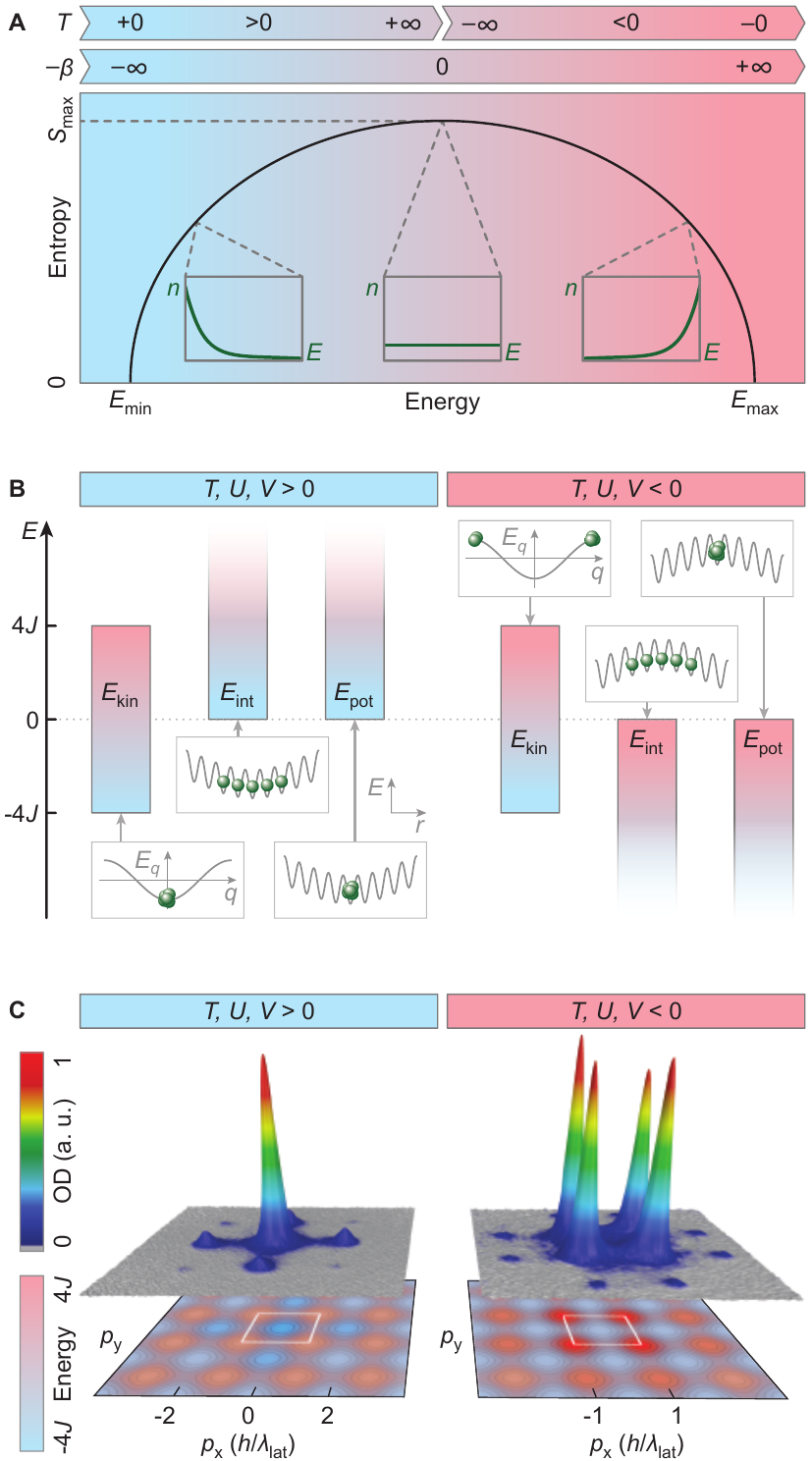} 
	\caption{{\small\textsf{Negative temperature in optical lattices. ({\bf A}) Sketch of entropy as a function of energy in a canonical ensemble possessing both lower ($E_{\mboxs{min}}$) and upper ($E_{\mboxs{max}}$) energy bounds. Insets: sample occupation distributions of single-particle states for positive, infinite, and negative temperature, assuming a weakly interacting ensemble. ({\bf B}) Energy bounds of the three terms of the two-dimensional (2D) Bose-Hubbard Hamiltonian: kinetic ($E_{\mboxs{kin}}$), interaction ($E_{\mboxs{int}}$), and potential ($E_{\mboxs{pot}}$) energy. ({\bf C}) Measured momentum distributions (TOF images) for positive (left) and negative (right) temperature states. Both images are averages of about 20 shots, both optical densities (OD) are individually scaled. The contour plots below show the tight-binding dispersion relation, momenta with large occupation are highlighted. The white square in the center indicates the first Brillouin zone.}}}
	\label{negative_T}
\end{figure}

\begin{figure*}[t]
	\centering
		\includegraphics[width=175mm]{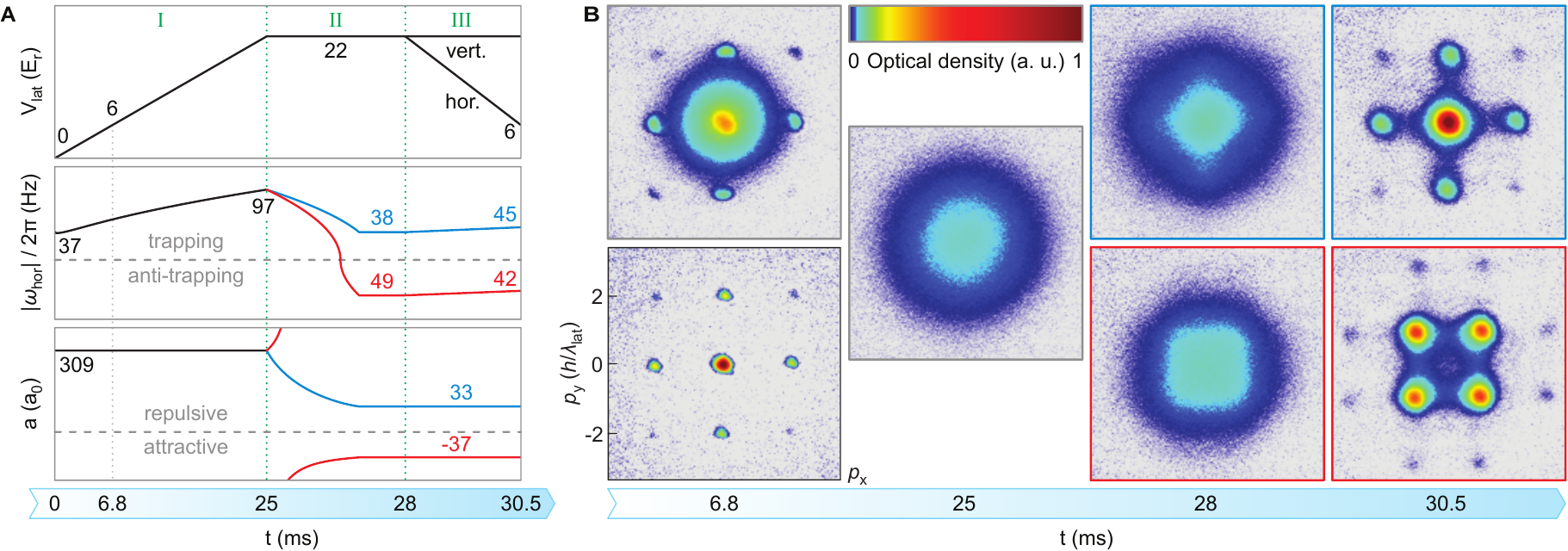} 
	\caption{{\small\textsf{Experimental sequence and TOF images. ({\bf A}) Top to bottom: lattice depth, horizontal trap frequency, and scattering length as a function of time. Blue indicates the sequence for positive, red for negative temperature of the final state. ({\bf B}) TOF images of the atomic cloud at various times $t$ in the sequence. Blue borders indicate positive, red negative temperatures. The initial picture in a shallow lattice at $t=6.8\,\mbox{ms}$ is taken once for a scattering length of $a=309(5)\,a_{\mboxs{0}}$ (top) as in the sequence, and once for $a=33(1)\,a_{\mboxs{0}}$ (bottom, OD rescaled by a factor of 0.25), comparable to the final images. All images are averages of about 20 individual shots. See also Fig.\ \ref{negative_T}C.}}}
	\label{sequence}
\end{figure*}

Absolute temperature $T$ is one of the central concepts of statistical mechanics and is a measure of e.g.\ the amount of disordered motion in a classical ideal gas. Therefore, nothing can be colder than $T=0$, where classical particles would be at rest. In a thermal state of such an ideal gas, the probability $P_i$ for a particle to occupy a state $i$ with kinetic energy $E_i$ is proportional to the Boltzmann factor,
\begin{equation}
\label{boltzmann}
P_i\propto e^{-E_i/k_{\mbox{\tiny{B}}}T},
\end{equation}
where $k_{\mboxs{B}}$ is Boltzmann's constant. An ensemble at positive temperature is described by an occupation distribution that decreases exponentially with energy. If we were to extend this formula to negative absolute temperatures, exponentially increasing distributions would result. Because the distribution needs to be normalizable, at positive temperatures a lower bound in energy is required, as the probabilities $P_i$ would diverge for $E_i\to -\infty$. Negative temperatures, on the other hand, demand an upper bound in energy \cite{Ramsey,Klein}. In daily life negative temperatures are absent, as kinetic energy in most systems, including particles in free space, only provides a lower energy bound. Even in lattice systems, where kinetic energy is split into distinct bands, implementing an upper energy bound for motional degrees of freedom is challenging, as potential and interaction energy need to be limited as well \cite{Mosk,Rosch}. So far, negative temperatures have been realized in localized spin systems \cite{Purcell,Oja,Ketterle}, where the finite, discrete spectrum naturally provides both lower and upper energy bounds. Here, we were able to realize a negative temperature state for motional degrees of freedom.

In Fig.\ \ref{negative_T}A we schematically show the relation between entropy $S$ and energy $E$ for a thermal system possessing both lower and upper energy bounds. Starting at minimum energy, where only the ground state is populated, an increase in energy leads to an occupation of a larger number of states and therefore an increase in entropy. As the temperature approaches infinity, all states become equally populated and the entropy reaches its maximum possible value $S_{\mboxs{max}}$. However, the energy can be increased even further if high-energy states are more populated than low-energy ones. In this regime the entropy decreases with energy, which, according to the thermodynamic definition of temperature \cite{Huang} ($1/T=\partial S/\partial E$), results in negative temperatures. The temperature is discontinuous at maximum entropy, jumping from positive to negative infinity. This is a consequence of the historic definition of temperature. A continuous and monotonically increasing temperature scale would be given by $-\beta=-1/k_{\mboxs{B}}T$, also emphasizing that negative temperature states are {\it hotter} than positive temperature states, i.e.\ in thermal contact heat would flow from a negative to a positive temperature system.

As negative temperature systems can absorb entropy while releasing energy, they give rise to several counterintuitive effects such as Carnot engines with an efficiency greater than unity \cite{Rosch}. Via a stability analysis for thermodynamic equilibrium we showed that negative temperature states of motional degrees of freedom necessarily possess negative pressure \cite{SM} and are thus of fundamental interest to the description of dark energy in cosmology, where negative pressure is required to account for the accelerating expansion of the universe \cite{Huterer}.

Cold atoms in optical lattices are an ideal system to create negative temperature states because of the isolation from the environment and independent control of all relevant parameters \cite{Zwerger}. Bosonic atoms in the lowest band of a sufficiently deep optical lattice are described by the Bose-Hubbard Hamiltonian \cite{Jaksch}
\begin{equation}
\label{hamiltonian}
H=-J\sum_{\langle i,j\rangle}\hat{b}_i^\dagger \hat{b}_j+\frac{U}{2}\sum_i \hat{n}_i(\hat{n}_i-1)+V\sum_i {\bf r}_i^2\hat{n}_i.
\end{equation}
Here, $J$ is the tunneling matrix element between neighboring lattice sites $\langle i,j\rangle$ and $\hat{b}_i$ and $\hat{b}_i^\dagger$ are the annihilation and creation operator, respectively, for a boson on site $i$, $U$ is the on-site interaction energy, $\hat{n}_i=\hat{b}_i^\dagger \hat{b}_i$ the local number operator, and $V\propto\omega^2$ describes the external harmonic confinement with ${\bf r}_i$ denoting the position of site $i$ with respect to the trap center and $\omega$ the trap frequency.

In Fig.\ \ref{negative_T}B we show how lower and upper bounds can be realized for the three terms in the Hubbard Hamiltonian. The restriction to a single band naturally provides upper and lower bounds for the kinetic energy $E_{\mboxs{kin}}$, but the interaction term $E_{\mboxs{int}}$ presents a challenge: because in principle all bosons could occupy the same lattice site, the interaction energy can diverge in the thermodynamic limit. For repulsive interactions ($U>0$), the interaction energy is only bounded from below but not from above, thereby limiting the system to positive temperatures; in contrast, for attractive interactions ($U<0$) only an upper bound for the interaction energy is established, rendering positive temperature ensembles unstable. The situation is different for the Fermi Hubbard model, where the Pauli principle enforces an upper limit on the interaction energy per atom of $U/2$ and thereby allows negative temperatures even in the repulsive case \cite{Aoki,Braun}. Similarly, a trapping potential $V>0$ only provides a lower bound for the potential energy $E_{\mboxs{pot}}$, while an anti-trapping potential $V<0$ creates an upper bound. Therefore, stable negative temperature states with bosons can exist only for attractive interactions and an anti-trapping potential.


In order to bridge the transition between positive and negative temperatures we used the $n=1$ Mott insulator \cite{Greiner} close to the atomic limit ($|U|/J\rightarrow\infty$), which can be approximated by a product of Fock states $|\Psi\rangle=\Pi_i \hat{b}_i^{\dagger}|0\rangle$. As this state is a many-body eigenstate in both the repulsive and the attractive case, it allows us to switch between these regimes, ideally without producing entropy. The employed sequence (Fig.\ \ref{sequence}A) is based on a proposal by Rapp et al.\ \cite{Rosch}, building on previous ideas by Mosk \cite{Mosk}. It essentially consists of loading a repulsively interacting Bose-Einstein condensate (BEC) into the deep Mott insulating regime (I in Fig.\ \ref{sequence}A), switching $U$ and $V$ to negative values (II), and finally melting the Mott insulator again by reducing $|U|/J$ (III). For comparison, we also created a final positive temperature state with an analog sequence.

The experiment started with a BEC of $1.1(2)\times 10^5$ $^{39}\mbox{K}$ atoms in a pure dipole trap with horizontal trap frequency $\omega_{\mboxs{dip}}$ ($V>0$) at positive temperature ($T>0$) and a scattering length of $a=309(5)\,a_{\mboxs{0}}$, with $a_{\mboxs{0}}$ the Bohr radius. We ramped up a three-dimensional optical lattice (I) with simple cubic symmetry to a depth of $V_{\mboxs{lat}}=22(1)\,E_{\mboxs{r}}$. Here $E_{\mboxs{r}}=h^2/(2m\lambda_{\mboxs{lat}}^2)$ is the recoil energy with Planck's constant $h$, the atomic mass $m$, and the lattice wavelength $\lambda_{\mboxs{lat}}=736.65\,\mbox{nm}$. The blue-detuned optical lattice provides an overall anti-trapping potential with a formally imaginary horizontal trap frequency $\omega_{\mboxs{lat}}$ that reduces the confinement of the dipole trap, giving an effective horizontal trap frequency $\omega_{\mboxs{hor}}=\sqrt{\omega_{\mboxs{dip}}^2+\omega_{\mboxs{lat}}^2}$. Once the atoms are in the deep Mott insulating regime where tunneling can essentially be neglected (tunneling time $\tau=h/(2\pi J)=10(2)\,\mbox{ms}$), we set the desired attractive (repulsive) interactions (II) to prepare a final negative (positive) temperature state using a Feshbach resonance \cite{Zaccanti}. Simultaneously, we decreased the horizontal confinement to an overall anti-trapping (trapping) potential by decreasing $\omega_{\mboxs{dip}}$. Subsequently, we decreased the horizontal lattice depths (III), yielding a final value of $U/J=-2.1(1)$ ($+1.9(1)$), and probed the resulting momentum distribution by absorption imaging after $7\,\mbox{ms}$ time-of-flight (TOF). The whole sequence was experimentally optimized to maximize the visibility of the final negative temperature state. We chose a 2D geometry for the final state in order to enable strong anti-trapping potentials and to avoid detrimental effects due to gravity \cite{SM}.

In Fig.\ \ref{sequence}B we show TOF images of the cloud for various times $t$ in the sequence, indicated in Fig.\ \ref{sequence}A. During the initial lattice ramp (at $V_{\mboxs{lat}}=6.1(1)E_{\mboxs{r}}$), interference peaks of the superfluid in the lattice can be observed ($t=6.8\,\mbox{ms}$ top). Because quantum depletion caused by the strong repulsive interactions already reduces the visibility of the interference peaks in this image \cite{Ketterle_depletion}, we also show the initial superfluid for identical lattice and dipole ramps, but at a scattering length of $a=33(1)\,a_{\mboxs{0}}$ ($t=6.8\,\mbox{ms}$ bottom). The interference peaks are lost as the Mott insulating regime is entered ($t=25\,\mbox{ms}$). In the deep lattice only weak nearest-neighbor correlations are expected, resulting in similar images for both repulsive and attractive interactions ($t=28\,\mbox{ms}$). After reducing the horizontal lattice depths back into the superfluid regime, the coherence of the atomic sample emerges again. For positive temperatures the final image at $t=30.5\,\mbox{ms}$ is comparable, albeit somewhat heated, to the initial one at $t=6.8\,\mbox{ms}$, whereas for attractive interactions sharp peaks show up in the corners of the first Brillouin zone, indicating macroscopic occupation of maximum kinetic energy. The spontaneous development of these sharp interference peaks is a striking signature of a stable negative temperature state for motional degrees of freedom. In principle, the system can enter the negative temperature regime following one of two routes: it either stays close to thermal equilibrium during the entire sequence or alternatively relaxes towards a thermal distribution during lattice ramp-down. Either way demonstrates the thermodynamic stability of this negative temperature state.


\begin{figure}[h!t]
	\centering
		\includegraphics[width=55mm]{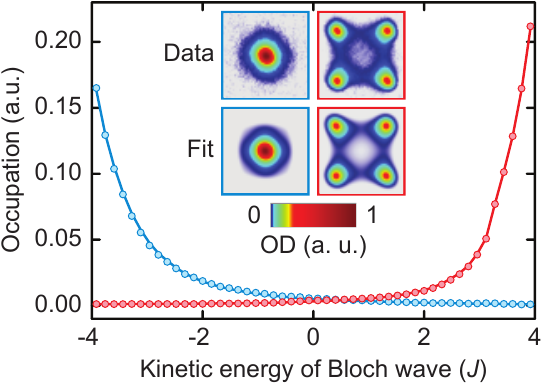}
		\caption{{\small\textsf{Occupation distributions. The occupation of the kinetic energies within the first Brillouin zone is plotted for the final positive (blue) and negative (red) temperature states. Points: experimental data extracted from band-mapped pictures. Solid lines: fits to a non-interacting Bose-Einstein distribution assuming a homogeneous system. Insets: top row: symmetrized positive (left) and negative (right) temperature images of the quasi-momentum distribution in the horizontal plane. Bottom row: fitted distributions for the two cases. Note that all distributions are broadened by the in situ cloud size \cite{SM}.}}}
	\label{distribution_function}
\end{figure}

In order to examine the degree of thermalization in the final states, we used band-mapped \cite{Jessen} images and extracted the kinetic energy distribution assuming a non-interacting lattice dispersion relation $E_{\mboxs{kin}}(q_{\mboxs{x}},q_{\mboxs{y}})$. The result is shown in Fig.\ \ref{distribution_function}, displaying very good agreement with a fitted Bose-Einstein distribution. The fitted temperatures of $T=-2.2J/k_{\mboxs{B}}$ and $T=2.7J/k_{\mboxs{B}}$ for the two cases only represent upper bounds for the absolute values $|T|$ of the average temperature because the fits neglect the inhomogeneous filling of the sample \cite{SM}. Both temperatures are slightly larger than the critical temperature $|T_{\mboxs{BKT}}|\approx 1.8J/k_{\mboxs{B}}$ \cite{Capogrosso2d} for the superfluid transition in an infinite 2D system but lie below the condensation temperature $|T_{\mboxs{C}}|=3.4(2)J/k_{\mboxs{B}}$ of non-interacting bosons in a 2D harmonic trap for the given average density \cite{SM}.

Ideally, entropy is produced during the sequence only in the superfluid/normal shell around the interim Mott insulator: while ramping to the deep lattice, the atoms in this shell localize to individual lattice sites and can subsequently be described as a $|T|=\infty$ system \cite{Braun}. Numerical calculations have shown that the total entropy produced in this process can be small \cite{Rosch}, as most of the atoms are located in the Mott insulating core. We attribute the observed additional heating during the sequence to non-adiabaticities during lattice ramp-down and residual double occupancies in the interim Mott insulator.


While the coherence length of the atomic sample can in principle be extracted from the interference pattern recorded after a long TOF \cite{Gerbier}, the experiment was limited to finite TOF, where the momentum distribution is convolved with the initial spatial distribution. By comparing the measured TOF images with theoretically expected distributions, we were able to extract a coherence length in the final negative temperature state of 3 to 5 lattice constants \cite{SM}.


\begin{figure}[h!t]
	\centering
		\includegraphics[width=84mm]{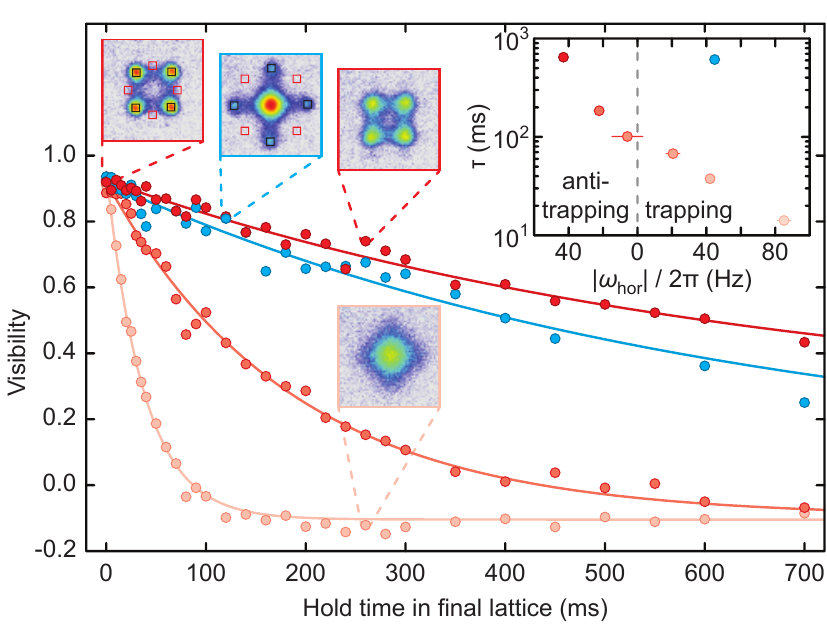} 
		\caption{{\small\textsf{Stability of the positive (blue) and negative (red) temperature states. Main figure: visibility ${\cal V}=(n_{\mboxs{b}}-n_{\mboxs{r}})/(n_{\mboxs{b}}+n_{\mboxs{r}})$ extracted from the atom numbers in the black ($n_{\mboxs{b}}$) and red ($n_{\mboxs{r}}$) boxes (indicated in the TOF images) plotted versus hold time in the final state for various horizontal trap frequencies. Dark red: $|\omega_{\mboxs{hor}}|/2\pi=43(1)\,\mbox{Hz}$ anti-trapping, medium red: $22(3)\,\mbox{Hz}$ anti-trapping, light red: $42(3)\,\mbox{Hz}$ trapping, blue: $45(3)\,\mbox{Hz}$ trapping. Inset: coherence lifetimes $\tau$ extracted from exponential fits (solid lines in main figure). The statistical error bars from the fits are smaller than the data points. The color scale of the images is identical to Fig.\ \ref{sequence}B (see also Fig.\ S3).}}}
	\label{lifetime}
\end{figure}

In order to demonstrate the stability of the observed negative temperature state, Fig.\ \ref{lifetime} shows the visibility of the interference pattern as a function of hold time in the final lattice. The resulting lifetime of the coherence in the final negative temperature state crucially depends on the horizontal trap frequencies (inset): lifetimes exceed $\tau=600\,\mbox{ms}$ for an optimally chosen anti-trapping potential, but an increasingly fast loss of coherence is visible for less anti-trapping geometries. In the case of trapping potentials, the ensemble can even return to metastable positive temperatures, giving rise to the small negative visibilities observed after longer hold times (Fig.\ S4). The loss of coherence probably originates from a mismatch between the attractive mean-field and the external potential, which acts as an effective potential and leads to fast dephasing between lattice sites.

The high stability of the negative temperature state for the optimally chosen anti-trapping potential indicates that the final chemical potential is matched throughout the sample such that no global redistribution of atoms is necessary. The remaining slow decay of coherence is not specific to the negative temperature state as we also observe comparable heating for the corresponding positive temperature case (blue data in Fig.\ \ref{lifetime} as well as the initial superfluid in the lattice. It probably originates from three-body losses and light-assisted collisions.

To summarize, we have created a negative temperature state for motional degrees of freedom. It exhibits coherence over several lattice sites, with coherence lifetimes exceeding $600\,\mbox{ms}$, and its quasi-momentum distribution can be reproduced by Bose-Einstein statistics at negative temperature. In contrast to metastable excited states \cite{Haller}, this isolated negative temperature ensemble is intrinsically stable and cannot decay into states at lower kinetic energies. It represents a stable bosonic ensemble at attractive interactions for arbitrary atom numbers; the negative temperature stabilizes the system against mean-field collapse that is driven by the negative pressure.


Negative temperature states can be exploited to investigate the Mott insulator transition \cite{Nagerl} as well as the renormalization of Hubbard parameters \cite{Will,Heinze} for attractive interactions. As the stability of the attractive gas relies on the bounded kinetic energy in the Hubbard model, it naturally allows a controlled study of the transition from stable to unstable by lowering the lattice depth, thereby connecting this regime with the study of collapsing BECs \cite{Wieman}, which is also of interest for cosmology \cite{Tatekawa}. Negative temperatures also significantly enhance the parameter space accessible for quantum simulations in optical lattices, as they enable the study of new many-body systems whenever the bands are not symmetric with respect to the inversion of kinetic energy. This is the case e.g.\ in triangular or Kagom\'e lattices, where in current implementations \cite{Stamper-Kurn} the interesting flat band is the highest of three sub-bands. In fermionic systems, negative temperatures enable e.g.\ the study of the attractive SU(3) model describing color superfluidity and trion (baryon) formation using repulsive $^{173}\mbox{Yb}$ \cite{Rapp}, where low losses and symmetric interactions are expected but magnetic Feshbach resonances are absent.

\vspace{\vspacelarge}

\begin{center}
{\large\textbf{Acknowledgments}}
\end{center}

\vspace{\vspacesmall}

\noindent
We thank Akos Rapp, Achim Rosch, Stephan Mandt, and Walter Hofstetter for helpful discussions and Daniel Garbe for technical assistance. We acknowledge financial support by the Deutsche Forschungsgemeinschaft (FOR801, Deutsch-Israelisches Kooperationsprojekt Quantum phases of ultracold atoms in optical lattices), the US Defense Advanced Research Projects Agency (Optical Lattice Emulator program), and Nanosystems Initiative Munich.

\renewcommand\refname{\center{\large{\textbf{References and Notes}}}}


%
%
%
%


\renewcommand{\theequation}{S\arabic{equation}}
\renewcommand{\thefigure}{S\arabic{figure}}
\setcounter{equation}{0}
\setcounter{figure}{0}

\vspace{0.5cm}

\newpage

\noindent
\begin{center}
{\textsf{\textbf{\huge{Supplementary materials}}}}
\end{center}

\vspace{0.2cm}

\noindent
\begin{center}
{\large\textbf{Negative pressure}}
\end{center}

\vspace{\vspacesmall}

In order to investigate the conditions for thermal equilibrium, we consider a gas in a box of volume $V_{\mboxs{box}}$ with fixed total energy $E$. In thermal equilibrium the entropy $S$ takes the maximal possible value under the given constraints. We note that the size of the box is only an upper limit for the volume $V$ of the gas, $V\leq V_{\mboxs{box}}$. In most cases this leads to the condition of a positive pressure $P>0$, forcing the gas to fill the whole box, $V=V_{\mboxs{box}}$ \cite{Lifshitz}. More generally, however, the maximum entropy principle only requires
\begin{equation}
\label{one}
\left.\frac{\partial S}{\partial V}\right|_E\geq0.
\end{equation}
If this derivative was negative, the system could spontaneously contract and thereby increase its entropy. As a consequence, there would be no equilibrium solution with $V=V_{\mboxs{box}}$ but instead the system would be unstable against collapse.

From the total differential of energy $dE=T\, dS-P\, dV$ one obtains
\begin{equation}
dS=\frac{1}{T}\, dE+\frac{P}{T}\,dV,
\end{equation}
which leads to
\begin{equation}
\left.\frac{\partial S}{\partial V}\right|_E=\frac{P}{T}.
\end{equation}
Thus absolute pressure and temperature necessarily have the same sign in equilibrium, $P/T\geq 0$, i.e.\ negative temperatures imply negative pressures and vice versa.

\label{som}

Since attractive interactions naturally lead to negative pressures, this illustrates why an attractively interacting BEC at positive temperatures is inherently unstable against collapse \cite{Ueda}. This is in contrast to the case studied here, where the negative temperature stabilizes the negative pressure system against collapse.

\vspace{\vspacelarge}

\noindent
\begin{center}
{\large\textbf{Experimental sequence}}
\end{center}

\vspace{\vspacesmall}

The sequence employed in this work is illustrated in Fig.\ 2A in the main text. We prepared a condensate of $N=105(14)\times 10^3$ $^{39}\mbox{K}$ atoms in a dipole trap with trapping frequencies of $\omega_{\mboxs{hor}}=2\pi\times 37(1)\,\mbox{Hz}$ and $\omega_{\mboxs{vert}}=2\pi\times 181(12)\,\mbox{Hz}$ along the horizontal and vertical directions, respectively. The horizontal trap frequency indicates the root mean square of the trap frequencies along the two horizontal directions. There is no detectable thermal fraction in TOF images of the condensed cloud. We created a Mott insulator by linearly ramping up a three-dimensional optical lattice (I in Fig.\ 2A) at a wavelength of $\lambda_{\mboxs{lat}}=736.65\,\mbox{nm}$ within $25\,\mbox{ms}$ to a lattice depth of $V_{\mboxs{lat}}\approx 22(1)\,E_{\mboxs{r}}$. The blue-detuned optical lattice provides an overall anti-trapping potential with a formally imaginary horizontal trap frequency $\omega_{\mboxs{lat}}$ that reduces the confinement of the dipole trap, giving an effective horizontal trap frequency $\omega_{\mboxs{hor}}=\sqrt{\omega_{\mboxs{dip}}^2+\omega_{\mboxs{lat}}^2}$. The scattering length during the lattice ramp was set to $a=309(5)\,a_{\mboxs{0}}$, resulting in a final interaction in the Mott insulating state of $U/J>800$. This loading sequence is designed to minimize doubly occupied sites in the Mott insulating state, as double occupancies promote atoms into higher bands when crossing the Feshbach resonance \cite{Busch}. During the lattice ramp, we increased the trapping frequencies to $\omega_{\mboxs{hor}}=2\pi\times 97(4)\,\mbox{Hz}$ and $\omega_{\mboxs{vert}}=2\pi\times 215(13)\,\mbox{Hz}$ by changing $\omega_{\mboxs{dip}}$, in order to increase the fraction of atoms in the Mott insulating core.

In the deep lattice, where tunneling can essentially be neglected (tunneling time $\tau=h/2\pi J=10(2)\,\mbox{ms}$), we ramped the scattering length within $2\,\mbox{ms}$ to its final value (II), either repulsive at $a=33(1)\,a_{\mboxs{0}}$ or attractive at $a=-37(1)\,a_{\mboxs{0}}$ ($|U/J|>80$) by crossing a Feshbach resonance at a magnetic field of 402.50(3)G \cite{Zaccanti}. At the same time, we decreased the horizontal confinement to $\omega_{\mboxs{hor}}=2\pi\times 38(4)\,\mbox{Hz}$ in the repulsive case or to a maximally anti-trapping potential with a formally imaginary trapping frequency of $|\omega_{\mboxs{hor}}|=2\pi\times 49(1)\,\mbox{Hz}$ in the attractive case by decreasing the power of the red-detuned dipole trap. The anti-confining potential is provided by the blue-detuned lattice beams. Subsequently, we linearly decreased the horizontal lattice depths (III) within $2.5\,\mbox{ms}$ to $V_{\mboxs{hor}}=6.1(1)\,E_{\mboxs{r}}$, but kept the vertical lattice at $V_{\mboxs{vert}}=22(1)\,E_{\mboxs{r}}$ to avoid effects due to gravity and to enable strong anti-trapping potentials (see below). During this lattice ramp, the horizontal trapping frequencies change only slightly. After instantaneously switching off all optical potentials, we ramped down the homogeneous magnetic field within $2\,\mbox{ms}$ and recorded absorption images along the vertical direction after a total TOF of $7\,\mbox{ms}$. The chosen TOF is a compromise between a sufficient transformation of momentum space into real space during TOF and a sufficiently large final optical density.

\vspace{\vspacelarge}

\noindent
\begin{center}
{\large\textbf{Inverting the external potential}}
\end{center}

\vspace{\vspacesmall}

In principle, negative temperature states can be created in three dimensions by additionally inverting the vertical confinement to an anti-trapping potential. In reality, however, inverting the vertical harmonic confinement without inverting gravity would also invert the gravitational sag, thereby creating a strong vertical gradient at the unchanged position of the atoms. While this issue could be mitigated by applying a suitable vertical magnetic field gradient, we instead chose to keep the vertical lattice strong. This has the additional advantage that more intense vertical lattice beams provide a stronger anti-trapping potential, enabling a total anti-trapping frequency of up to $|\omega_{\mboxs{hor}}|=2\pi\times 43(1)\,\mbox{Hz}$.

For an exact inversion of the interaction and potential energy terms in the Hamiltonian, the horizontal confinement should have been precisely inverted to an anti-trapping potential of frequency $|\omega_{\mboxs{hor,f}}|=\omega_{\mboxs{hor,i}}=2\pi\times 97\,\mbox{Hz}$ (f: final, i: initial) simultaneously with a Feshbach ramp to a negative scattering length of $a_{\mboxs{f}}=-a_{\mboxs{i}}=-309\,a_{\mboxs{0}}$ and the sequence thereafter should have mirrored the loading sequence. In the experiment, however, completely inverting the potential would only be possible with an additional blue-detuned anti-trapping beam. To compensate the resulting mismatch in potential energy, we also decreased the interaction strength simultaneously to the harmonic confinement such that the ratio $|U|/|\omega_{\mboxs{hor}}|^2$ remains approximately constant. If we assume that the density distribution before entering the Mott insulator is in global thermal equilibrium and density is hardly redistributed in the deep lattice, the sum of potential and mean-field interaction energy should again be approximately constant throughout the cloud after ramping back into the superfluid regime. This minimizes the effective potential for the atoms and the associated dephasing in the final state. We note that ramping down the lattice depth within $25\,\mbox{ms}$, as in the initial loading sequence, would result in a comparable final visibility ($4\%$ lower than for $2.5\,\mbox{ms}$) for an anti-trapping potential with $|\omega_{\mboxs{hor}}|=2\pi\times 43(1)\,\mbox{Hz}$. The shown lifetime measurements over a large range of $|\omega_{\mboxs{hor}}|$, however, are only possible with a fast ramp, where dephasing during the ramp due to strong effective potentials is less important.

\vspace{\vspacelarge}

\noindent
\begin{center}
{\large\textbf{Quasi-momentum distribution}}
\end{center}

\vspace{\vspacesmall}

In order to measure the quasi-momentum distribution of the final states we used a band-mapping technique \cite{Jessen}, i.e.\ we linearly ramped down the final lattice in $60\,\mu\mbox{s}$ followed by $7\,\mbox{ms}$ TOF and averaged about 20 images for both the final negative and positive temperature states. We note that these images do not directly represent the quasi-momentum distribution of the ensembles, but are convolved with the in situ density distribution. As the lattice axes along the horizontal directions are not perfectly orthogonal, we rectified the slight asymmetry in the position of the coherence peaks: We fitted the four main peak positions of a negative temperature ensemble and, by applying a shearing transformation, mapped them onto the corners of a square. For a reliable fit of 2D Bose-Einstein distributions (see below) we also equalized the varying heights of the four peaks by multiplication with a linearly interpolated normalization map. For the positive temperature state, we extracted the necessary shearing transformation from the positions of the first order coherence peaks in an image without band-mapping. We did not rescale the height of the single coherence peak in the band-mapped image. The resulting images are shown as the upper insets in Fig.\ 3 in the main text.

After symmetrizing the data, we assigned horizontal quasi-momenta $q_{\mboxs{x}}$ and $q_{\mboxs{y}}$ to the individual pixels of the images, and extracted the quasi-momentum distribution $n(q_{\mboxs{x}},q_{\mboxs{y}})$ in the first Brillouin zone, which is still convolved with the initial density. From this, we obtained the kinetic energy distribution, i.e. the number of atoms with a given kinetic energy, by use of the tight-binding dispersion relation at the given lattice depth,
\begin{equation}
\label{dispersion_relation}
E_{\mboxs{kin}}(q_{\mboxs{x}},q_{\mboxs{y}})=-2J\left[\cos\left(q_{\mboxs{x}}d\right)+\cos\left(q_{\mboxs{y}}d\right)\right],
\end{equation}
where $d=\lambda_{\mboxs{lat}}/2$ is the lattice constant. This distribution was then normalized by the density of states in the 2D optical lattice in order to obtain the occupation $\rho_{\mboxs{exp}}(E_{\mboxs{kin}})$ per Bloch wave as a function of energy. The result is shown as the blue and red data points in Fig.\ 3 in the main text for the final positive and negative temperature states, respectively.

\vspace{\vspacelarge}

\noindent
\begin{center}
{\large\textbf{Bose-Einstein fits}}
\end{center}

\vspace{\vspacesmall}

In order to extract the temperature of the final states, we fitted the quasi-momentum distributions, as detailed above, by a Bose-Einstein distribution function for the kinetic energy,
\begin{equation}
\label{be_statistics}
n(q_{\mboxs{x}},q_{\mboxs{y}})=\frac{1}{e^{(E_{\mboxs{kin}}(q_{\mboxs{x}},q_{\mboxs{y}})-\mu)/k_BT}-1}+o,
\end{equation}
which was convolved with a Gaussian (see below). Independent fitting parameters include the chemical potential $\mu$, the temperature $T$ and a constant offset $o$. The results of these fits are shown as the lower insets in Fig.\ 3 in the main text. From the fits, the occupation $\rho_{\mboxs{fit}}(E_{\mboxs{kin}})$ of the Bloch waves as a function of energy was extracted analogously to the experimental data (see above), and is shown as solid curves in Fig.\ 3. 

The very good reproduction of the data with a Bose-Einstein distribution function, together with the appearance of stable coherence peaks, indicates that the final negative and positive temperature states are thermalized. This fitting procedure, however, neglects interaction effects and potential energy and overestimates the filling by assuming a homogeneous density of one atom per site (see below), while the experiment uses an inhomogeneous system with unity filling only in the center. Therefore, the fitted temperatures of $2.7J/k_{\mboxs{B}}$ and $-2.2J/k_{\mboxs{B}}$ are systematically too large, considering absolute values, and represent only upper bounds.

During the final lattice ramp of $2.5\,\mbox{ms}$ from the Mott insulating to the superfluid regime, not more than 2 integrated tunneling events can take place, limiting the overall mass transport during ramp-down. To a good approximation, we can therefore estimate the final average filling $\bar{n}$ in the superfluid regime by $\bar{n}$ in the Mott insulating state. The average filling in the atomic limit depends on the entropy of the cloud, and can range for our parameters from $\bar{n}=1$ in the zero entropy case to $\bar{n}\approx0.7$ for $S/N\approx1.0\,k_{\mboxs{B}}$ and $\bar{n}\approx0.5$ for $S/N\approx1.5\,k_{\mboxs{B}}$. Performing the above fit for average fillings of $\bar{n}=0.7$ and $\bar{n}=0.5$ yields temperatures that are $20(3)\%$ and $33(5)\%$, respectively, below the values reported above.

\vspace{0.2cm}

\noindent
{\textbf{Fitting details:}}

\vspace{\vspacesmall}

In the fit, we convolved the Bose-Einstein distribution function of eq.\ \ref{be_statistics} with an elliptical Gaussian in order to account for both the convolution with the initial density distribution as well as the vertical extension of the cloud after TOF. The ellipticity appears because the vertical lattice axis is not perfectly parallel to the imaging axis. We fixed both aspect ratio and angle of the elliptical Gaussian to values obtained from separate fits of the central peak in a positive temperature image. Therefore only the width $\sigma_{\mboxs{G}}$ remains as an additional fitting parameter for the Bose-Einstein fit, which contains in total five fitting parameters: $\mu$, $T$, $o$, $\sigma_{\mboxs{G}}$, and the size $l_{\mboxs{BZ}}$ of the first Brillouin zone after TOF. The experimental data was normalized to the number of quasi-momentum states used in the fit, i.e.\ the fit corresponds to unity filling.

While monitoring the sum of squared residuals showed a good stability of the fit for negative temperatures, it is not possible to use $\sigma_{\mboxs{G}}$ as free parameter in the positive temperature case, because $T$ and $\sigma_{\mboxs{G}}$ are not independent when fitting only a single peak. Assuming identical convolution functions for negative and positive temperatures, we instead fixed $\sigma_{\mboxs{G}}$ to the value obtained from fitting the negative temperature image. Similarly, as $l_{\mboxs{BZ}}$ cannot be obtained from a single peak, it was fixed to the fitted distance of the first order coherence peaks in a positive temperature image without band-mapping. The positive temperature fitting routine therefore contains only $\mu$, $T$, and $o$ as free parameters.

\vspace{\vspacelarge}

\noindent
\begin{center}
{\large\textbf{Critical temperature}}
\end{center}

\vspace{\vspacesmall}

While, contrary to the three-dimensional case, there is no condensation into a Bose-Einstein condensate (BEC) at ${\bf p}=0$ in 2D free space \cite{Hohenberg}, condensation is nonetheless possible for non-interacting bosons in a 2D harmonic trap \cite{Bagnato, Mullin}. Here the critical temperature is given by
\begin{equation}
\label{T_c}
T_{\mboxs{C}}=\sqrt{\frac{6N_{\mboxs{2D}}}{\pi^2}}\frac{\hbar\omega_{\mboxs{hor}}}{k_{\mboxs{B}}},
\end{equation}
where $N_{\mboxs{2D}}$ denotes the atom number.

This formula can be adapted to the lattice case via the effective mass approximation for small quasi-momenta \cite{Morsch},
\begin{equation}
-2J[\cos(q_{\mboxs{x}}d)+\cos(q_{\mboxs{y}}d)]\approx-2J+\frac{\hbar^2(q_{\mboxs{x}}^2+q_{\mboxs{y}}^2)}{2m_{\mboxs{eff}}},
\end{equation}
with an effective mass of
\begin{equation}
m_{\mboxs{eff}}=\frac{\hbar^2}{2Jd^2}.
\end{equation}
For the employed lattice depth of $V_{\mboxs{lat}}=6\,E_{\mboxs{r}}$ we obtain $m_{\mboxs{eff}}/m=2.008\approx2$ and can derive the effective trap frequency in the lattice as
\begin{equation}
\omega_{\mboxs{hor}}^{\mboxs{eff}}=\sqrt{\frac{m}{m_{\mboxs{eff}}}}\,\omega_{\mboxs{hor}},
\end{equation}
which can be inserted for the bare trap frequency in formula \ref{T_c}.

The atom number in the central layer can be estimated by assuming the initial $n=1$ Mott insulator in the atomic limit to form an ellipsoid of volume $V=\frac{4}{3}\pi R^3/\gamma=d^3 N$, with the radius $R$ in the horizontal directions and the aspect ratio $\gamma=\omega_{\mboxs{vert}}/\omega_{\mboxs{hor}}$ of the trap. The area of the central layer is given by $A=\pi R^2=\pi\left(3\gamma\,d^3 N/4\pi\right)^{2/3}=d^2N_{\mboxs{2D}}$, leading to
\begin{equation}
N_{\mboxs{2D}}=\pi\left(\frac{3\gamma N}{4\pi}\right)^\frac{2}{3}\approx 4.6(4)\times 10^3.
\end{equation}

For our parameters, this yields a critical temperature of $T_{\mboxs{C}}=3.4(2)J/k{\mboxs{B}}$. One expects a quasi-condensate with fluctuating phase below $T_{\mboxs{C}}$ which only well below $T_{\mboxs{C}}$ turns into a true condensate with small fluctuations on distances comparable to the Thomas-Fermi size \cite{Petrov}. We note, however, that this theory does not hold for the interacting case in the thermodynamic limit, where the BEC transition is replaced by a Berezinskii-Kosterlitz-Thouless (BKT) transition into a superfluid \cite{Dalibard}, which is expected to occur at $T_{\mboxs{C}}\approx 1.8 J/k_{\mboxs{B}}$ \cite{Capogrosso2d} for the parameters used here. Nonetheless, Monte-Carlo calculations have shown that quasi-condensate correlations appear well above the critical temperature for the BKT transition \cite{Svistunov}.

\vspace{\vspacelarge}

\noindent
\begin{center}
{\large\textbf{Extraction of coherence length}}
\end{center}

\vspace{\vspacesmall}

In principle, the coherence length $l_{\mboxs{c}}$ in the final negative temperature state can be directly extracted from the peak width in the interference pattern recorded after long TOF. After finite TOF, however, the momentum distribution of the sample is convolved with the initial spatial distribution of the atoms. Following \cite{Gerbier}, we therefore model the one-dimensional (1D) distribution after finite TOF by $n(k)\propto|\tilde{w}_0(k)|^2S(k)$, where the envelope function $\tilde{w}_0(k)$ is the Fourier transform of the on-site Wannier function and $k=p/\hbar$. The interference term
\begin{equation}
S(k)=\sum_{r_\mu,r_\nu}e^{ik(r_\mu-r_\nu)-i\frac{m}{2\hbar t}(r_\mu^2-r_\nu^2)-\frac{r_\mu^2+r_\nu^2}{4R^2}-\frac{|r_\mu-r_\nu|}{l_{\mboxs{c}}}}
\end{equation}
assumes an initial Gaussian density distribution with the standard deviation $R=38(2)d$ being the radius of the central 1D system (see above), and a phase coherence that decays exponentially over $l_{\mboxs{c}}$. The coordinate of lattice site $\mu$ is indicated by $r_{\mu}$.

\begin{figure}[h!t]
	\centering
		\includegraphics[width=84mm]{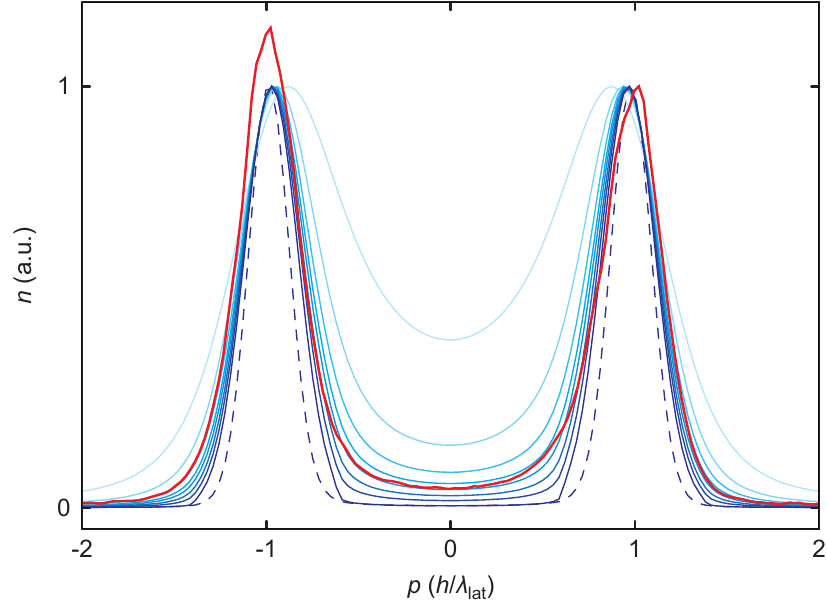} 
		\caption{{\small\textsf{Extraction of coherence length from background. Red curve: measured optical density along a cut through two peaks in the TOF image ($t=7\,\mbox{ms}$). It is scaled to a value of 1 for the lower peak. Blue curves: normalized densities extracted from one-dimensional calculations assuming $R=38d$ (dashed curve: $R=28d$) and, from light to dark, $l_{\mboxs{c}}=(1, 2, 3, 4, 5, 7, 11, 38)d$ (dashed curve: $l_{\mboxs{c}}=28d$).}}}
	\label{coherence_length}
\end{figure}

\begin{figure}[h!t]
	\centering
		\includegraphics[width=84mm]{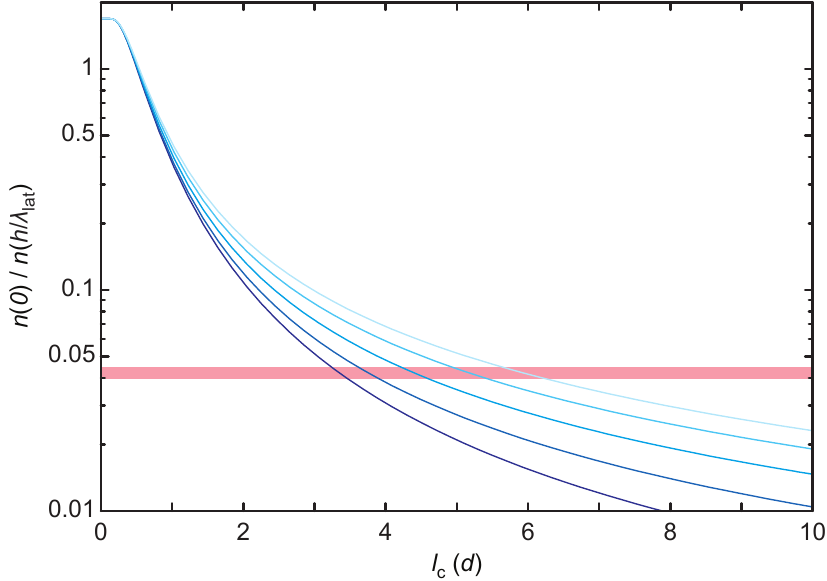} 
		\caption{{\small\textsf{Extraction of coherence length from peak density and background. Blue curves: the calculated value $n(0)/n(h/\lambda_{\mboxs{lat}})$ for a given TOF is plotted versus $l_{\mboxs{c}}$ for various radii, from light to dark, of $R=(8, 18, 28, 38, 48)d$. Red-shaded area: value extracted from the experimental data for the range of measured peak densities.}}}
	\label{coherence_length_2}
\end{figure}

In Fig. \ref{coherence_length} we show calculated interference patterns for various $l_{\mboxs{c}}$ together with experimental data. While the observed peak width is consistent with phase coherence over the whole sample, there are two subtleties to consider: first, for large $l_{\mboxs{c}}$ the calculated peak width depends critically on $R$, and second, the above 1D calculation neglects the other dimensions. While these could, in a first approximation, be included by averaging over various 1D systems with different $R$, the calculation nonetheless depends crucially on the initial in-situ distribution. The contrast of the 1D interference signal, on the other hand, depends less strongly on $R$ (see Fig. \ref{coherence_length_2}) and is therefore also less sensitive to averaging over several systems. From this signal we can deduce a lower bound for $l_{\mboxs{c}}$ of 3 to 5 lattice sites. Atoms from the interim Mott insulating core carry little entropy and are expected to evolve into a superfluid for sufficiently slow ramps. On the contrary, atoms from the former superfluid/normal shell and from double occupancies carry a lot of entropy and will show up as an overall $|T|=\infty$ Gaussian background in TOF images, i.e.\ they are also expected in the region between the interference peaks. This simple model therefore systematically underestimates $l_{\mboxs{c}}$ in the central region.

\vspace{\vspacelarge}

\noindent
\begin{center}
{\large\textbf{Lifetime of coherence}}
\end{center}

\vspace{\vspacesmall}

\begin{figure}[h!t]
	\centering
		\includegraphics[width=84mm]{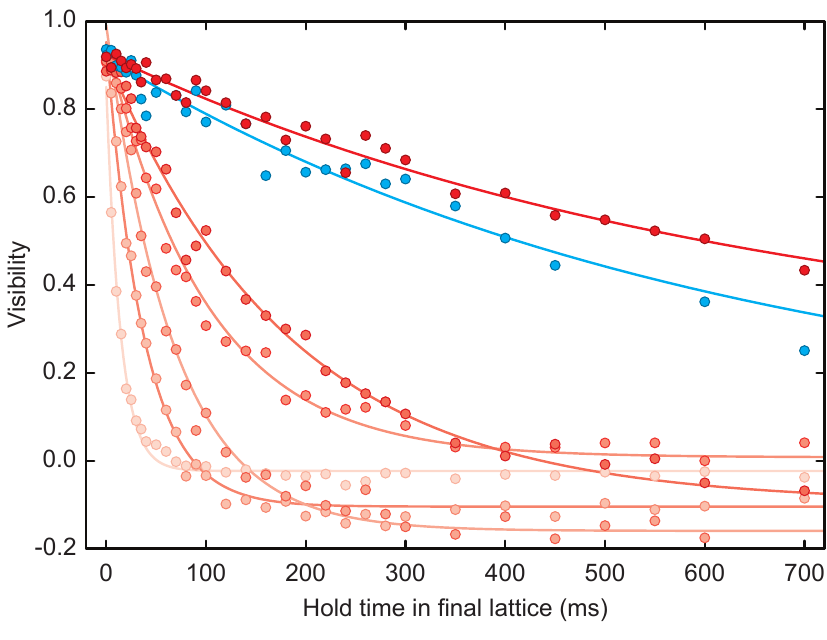} 
		\caption{{\small\textsf{Stability of the final states. Data: visibility plotted versus hold time in the final state for various horizontal trap frequencies. Red: attractive interactions, from dark to light, $|\omega_{\mboxs{hor}}|/2\pi=(43(1), 22(3), 6(9))\,\mbox{Hz}$ anti-trapping, $(21(4), 42(3), 85(4))\,\mbox{Hz}$ trapping, blue: repulsive interactions, $45(3)\,\mbox{Hz}$ trapping. Solid lines: exponential fits.}}}
	\label{lifetime_SOM_2}
\end{figure}

\begin{figure*}[h!b]
	\centering
		\includegraphics[width=175mm]{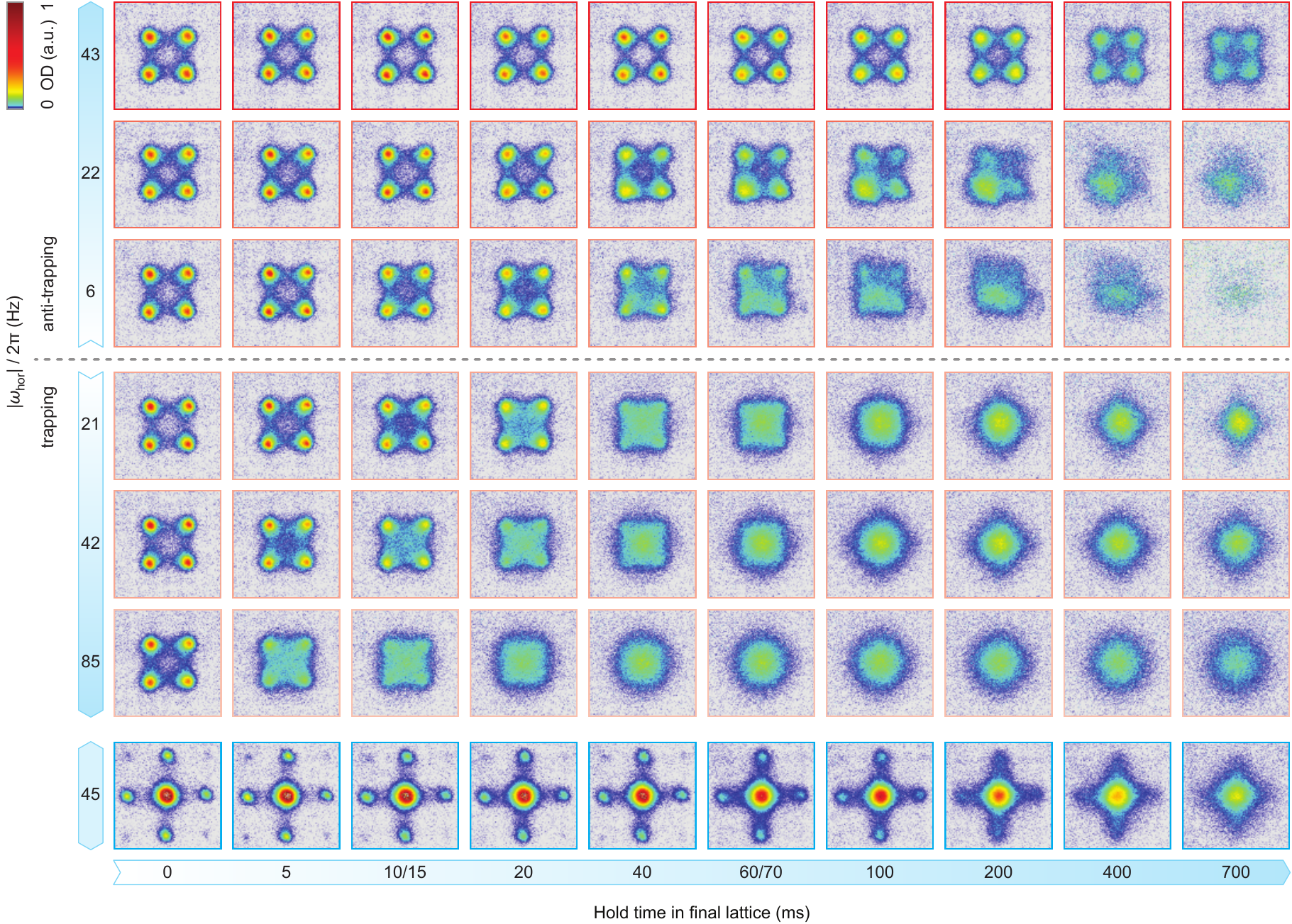} 
		\caption{{\small\textsf{Sample images of the stability measurement. Shown are TOF images for various horizontal trap frequencies $|\omega_{\mboxs{hor}}|$ for negative (red borders) and positive (blue borders) temperatures after variable hold times in the final lattice, corresponding to the data in Fig.\ 4 in the main text and in Fig.\ \ref{lifetime_SOM_2}.}}}
	\label{lifetime_SOM}
\end{figure*}

In Fig.\ \ref{lifetime_SOM_2} we present the complete data and exponential fits leading to the coherence lifetimes plotted in the inset of Fig.\ 4 in the main text, and in Fig.\ \ref{lifetime_SOM} we show corresponding sample images. We observe a high stability of the negative temperature interference pattern for the optimal anti-trapping potential with a frequency of $|\omega_{\mboxs{hor}}|=2\pi\times 43(1)\,\mbox{Hz}$, and also for the positive temperature pattern for the analogue trapping potential with frequency $|\omega_{\mboxs{hor}}|=2\pi\times 45(3)\,\mbox{Hz}$. For weaker anti-trapping potentials ($|\omega_{\mboxs{hor}}|=2\pi\times 22(3)\,\mbox{Hz}$ and $2\pi\times 6(9)\,\mbox{Hz}$), we not only observe shorter lifetimes of the coherence, but also distortions of the cloud from residual non-harmonic terms in the potentials. The observed reduction of total atom number which is strongest for the weakest external potentials can be explained by atoms leaving the trap in the horizontal plane. For trapping (i.e.\ not anti-trapping) potentials, where the lifetime is reduced even further, faint cross-like structures appear for long hold times, indicating that the system can return to large metastable positive temperatures.


\vspace{\vspacereferences}

\renewcommand\refname{\center{\large{

\end{document}